\documentclass{article}
\usepackage{waspaa19,amsmath,graphicx,url,times}
\usepackage{color}
\usepackage{multicol,multirow}
\usepackage{todonotes}

\title{CITY CLASSIFICATION FROM MULTIPLE REAL-WORLD SOUND SCENES}

\name{Helen L. Bear$^{1}$,\thanks{This work was conducted whilst HLB was on secondment to Tampere University. HLB is funded under EPSRC grant EP/R01891X/1. EB is supported by RAEng Research Fellowship RF/128 and a Turing Fellowship. This work was supported by the European Research Council under the ERC Grant Agreement 637422 EVERYSOUND.} 
      Toni Heittola$^{2}$, 
      Annamaria Mesaros$^{2}$, 
      Emmanouil Benetos$^{1,3}$,
      Tuomas Virtanen$^{2}$}
\address{$^1$ School of EECS, Queen Mary University of London, UK \ \ \ \ $^3$ The Alan Turing Institute, UK\\ 
         $^2$ Signal Processing Department, Tampere University, Finland\\
          \{h.bear, emmanouil.benetos\}@qmul.ac.uk \ \{toni.heittola, annamaria.mesaros, tuomas.virtanen\}@tut.fi\\
}

\begin{document}

\ninept
\maketitle

\begin{sloppy}

\begin{abstract}
The majority of sound scene analysis work focuses on one of two clearly defined tasks: acoustic scene classification or sound event detection. Whilst this separation of tasks is useful for problem definition, they inherently ignore some subtleties of the real-world, in particular how humans vary in how they describe a scene. Some will describe the weather and features within it, others will use a holistic descriptor like `park', and others still will use unique identifiers such as cities or names. 
In this paper, we undertake the task of automatic city classification to ask whether we can recognize a city from a set of sound scenes? In this problem each city has recordings from multiple scenes. 
We test a series of methods for this novel task and show that a simple convolutional neural network (CNN) can achieve accuracy of 50\%.  This is less than the acoustic scene classification task baseline in the DCASE 2018 ASC challenge on the same data. A simple adaptation to the class labels of pairing city labels with grouped scenes, accuracy increases to 52\%, closer to the simpler scene classification task. Finally we also formulate the problem in a multi-task learning framework and achieve an accuracy of 56\%, outperforming the aforementioned approaches.
\end{abstract}

\begin{keywords}
Acoustic scene classification, location identification, city classification, computational sound scene analysis.
\end{keywords}

\section{Introduction}
The majority of sound scene analysis research focuses on either acoustic scene classification (ASC) or sound event detection (SED)~\cite{virtanen2018computational}. This task separation is useful for clear problem definition, but it ignores some subtleties of the real-world, such as how humans vary in how they describe a scene. Some will describe the weather and features within it, others will use a holistic descriptor like `park', and others still will use unique identifiers such as cities or names~\cite{Guastavino2018}. In this work we undertake automatic city classification to predict a city from a set of sound scenes. In this problem each city has recordings from multiple scenes.

City classification from soundscapes is a task with limited prior work. We define it as the task of correctly identifying a city from different acoustic scenes across the city. 
Preliminary audio geotagging work used sound scenes augmented with city-inspecific sound events \cite{7545041,kumar2017audio}. They supplement augmented sound scenes with sound event models and rather than predicting a city label, the problem is modelled as a retrieval task and performance is measured by recalling audio for the most similar city. 

With real-world sound scenes as inputs, city classification is an interesting task because of its applications. As a practical example, precise geo-location from the background sounds of an emergency 999 call would assist the response times~\cite{1416352}. Furthermore, humans do not always describe scenes based on a set list of categories e.g.~park, street, but rather by what they recognise about a scene such as a well known place e.g. Edinburgh~\cite{4036742}. As such, using city classes as labels expands the scope of current sound scene analysis work to become more like an open set problem as motivated in \cite{bear2018extensible} because it expands the range of possible predictions to anywhere on earth 
and more akin to human-based descriptions, without addressing out-of-vocabulary labels. Therefore in this work, we use an existing dataset which contains geographic labels (these labels have not been used before) and carry out a set of tests to determine which is most useful for city classification.

In an open set taxonomy the relationships between labels and sets of labels varies. Whilst some hierarchies may exist, in other cases labels are simply constructs of human observations on the world \cite{bear2018extensible}. In this case there is a many-to-many relationship between scene and city labels, that is many types of scene can appear in a single city and multiple cities will contain each scene.

We suspect that as there is greater variation within the city data due to scene specific similarities in sounds, city classification is a harder problem. We will demonstrate this with experiments including multi-class, multi-label, and multi-task classification, with the latter achieving greatest accuracy.
The rest of this paper is organised as follows: we summarise the dataset and our feature extraction process in Section~\ref{sec:data}, before describing the classification methods in Section~\ref{sec:method} and analysing the results in Section~\ref{sec:results}. We conclude in Section~\ref{sec:conclusions} with the lessons learned for this new problem.  
\vspace{-0.3em}
\section{Data}
\label{sec:data}
We use audio data from the DCASE 2018 Acoustic Scene Classification (ASC) subtask 1A \cite{Mesaros2018_DCASE}. This dataset consists of recordings from six cities: Barcelona, Helsinki, London, Paris, Stockholm, and Vienna, and is partitioned so that the training subset contains for each city approximately $70\%$ of recording cities (balanced by city). It was originally labelled for ten sound scenes and recorded in ten predefined locations~\cite{Mesaros2018_DCASE} at random times between 9am and 9pm on different days of the week. Of the total development segments released (8640), 6122 segments are included in the training subset and 2518 segments in the evaluation subset. We use the evaluation subset as validation and our test set is another 2518 test files previously held back to review the challenge submissions. We maintain these same divisions for all our experiments and have ensured that inter-class ratios are approximately the same for all splits. Each acoustic scene has 864 segments (144 minutes of audio). The dataset contains in total 24 hours of audio. The original recordings were split into segments with a length of 10 seconds. The equipment used for recording consists of a binaural Soundman OKM II Klassik/studio A3 electret in-ear microphone and a Zoom F8 audio recorder using 48 kHz sampling rate and 24 bit resolution.

\begin{table}[t]
    \centering
    \caption{Distances in miles between each pair of cities in the DCASE 2018 ASC dataset and Euclidean distance between the mean features for each city in the training data.}
    \begin{tabular}{l|r|r}
    City from and to & Miles & Feature Euclidean   \\
    \hline \hline
    London-Paris &  234 & 0.67\\
    Helsinki-Stockholm & 302 & 4.34 \\
    Barcelona-Paris & 620 & 3.60 \\
    Paris-Vienna & 721 & 2.36 \\
    Barcelona-London  & 829 & 3.31 \\
    London-Vienna & 867 & 2.16 \\
    Stockholm-Vienna & 876 & 6.11\\
    Helsinki-Vienna & 1007 & 2.08\\
    Barcelona-Vienna & 1034 & 4.93 \\
    London-Stockholm & 1083 & 4.48 \\ 
    Paris-Stockholm & 1094 & 4.79\\
    Helsinki-London & 1384 & 1.35 \\
    Helsinki-Paris & 1395 & 1.88\\
    Barcelona-Stockholm & 1642 & 2.04\\
    Barcelona-Helsinki & 1897 & 3.20\\
    \end{tabular}
    \label{tab:citydistances}
\end{table}
Features are extracted for every recording as the log of mel-spectrogram energies using 128 mel bands, hop length of 512, and an STFT size of 2048. For every data sample we subtract the mean of the training data and divide by the standard deviation of the whole training data per frequency bin. Finally, each sound scene feature is smoothed over time (25 frames) as per \cite{bisot2017feature}, as this has been shown to be effective in ASC.
Table~\ref{tab:citydistances} lists the Euclidean distance (ED) between the mean training data features for each pair of cities. These are ordered by geographic distances in miles and there is no reported correlation but there is a big variation from London-Paris with $ED=0.67$ compared to Stockholm-Vienna with $ED=6.11$. We later use these to interpret city confusions. 
\vspace{-0.3em}
\section{Methods}
\label{sec:method}
With our prepared data we undertake a series of classification tests to experimentally review city classification. These include: validating our model against the DCASE ASC task benchmark, a traditional six-class classification task, using the scene labels as prior information, pairing scenes and city labels, grouping scenes, multi-label classification with two labels per recording, and finally, multi-task learning. All methods use the same train/dev/test data split, $200$ epochs of training, and the same model unless specific alterations are required and described. All code is available online\footnote{\url{github.com/drylbear/soundscapeCityClassification}}.

\begin{table}[t]
    \centering
    \caption{CNN structure and parameters. Adam optimiser with $LR=0.001$, $beta_1=0.9$, $beta_2=0.999$, $epsilon=None$, $decay=0.0$, $amsgrad=False$.}
    \resizebox{\columnwidth}{!}{%
    \begin{tabular}{r|l}
    \hline
    Layer & params \\
    \hline \hline
    Convolutional & filters=32, kernel=(7,7)   \\
    BatchNormalization & \\
    MaxPooling2D & pool\_size=(5,5), strides=2, padding=`same' \\
    Dropout & prob\_drop\_conv=0.3 \\
    
    Convolutional & filters=64,  kernel=(7,7)   \\
    BatchNormalization & \\
    MaxPooling2D & pool\_size=(4,7), strides=2, padding=`same' \\
    Dropout & prob\_drop\_conv=0.3 \\
    
    Convolutional & filters=128, kernel=(2,2)   \\
    BatchNormalization & \\
    MaxPooling2D & pool\_size=(5,5), strides=2, padding=`same' \\
    Dropout & prob\_drop\_conv=0.3 \\
    
    Flatten & \\ 
    Dense & filters=64, activation=`relu' \\
    Dropout & prob\_drop\_hidden=0.3 \\
    BatchNormalization & \\
    
    Dense & filters=6, activation=`softmax' \\
     \hline
    \end{tabular}}%
    \label{tab:params}
\end{table}

This work begins with the benchmark ASC CNN model from the DCASE 2018 ASC subtask 1A challenge \cite{Mesaros2018_DCASE}. This model scores $59.7\%$ accuracy over ten scene classes. We refine the model with minor tweaks such as simply adding one extra convolutional layer (line three in Table~\ref{tab:params}) to increase scene accuracy on the same train/dev/test data arrangement. We test other model options such as including LSTM layers but these produce inaccurate city predictions (one LSTM layer reduces accuracy to $21\%$) so they are not pursued. Table~\ref{tab:params} describes the final model architecture.

\subsection{Experiments}
\paragraph*{Multi-class city classification}
This is a simple six-class, single-label multi-class problem. Using the development dataset available from the DCASE 2018 ASC task, which includes labels for cities, we build a simple CNN for the six European cities: Barcelona, Helsinki, London, Paris, Stockholm, Vienna. For each city there are recordings from all ten scenes. This gives us a significantly greater variation within each city class than in traditional ASC tasks. 
\vspace{-1.35em}
\paragraph*{Scene specific city prediction}
Next, we train ten scene-specific city classifiers. In this case, each set of city predictions is specific to the scene used for training. There is less training data for each model.
\vspace{-1.35em}
\paragraph*{Pairing scenes and city labels}
In the next three methods, we test altering the labelling scheme with the hope to better understand the confusions between classes. This is considered particularly useful given the relationship between city and scene labels on the same data. The first of these three is a 60-class CNN where each class label is denoted $[scene\_city]$. 
\vspace{-1.35em}
\paragraph*{Grouped scenes}
Given the possible confusions between similar scenes, we run a second relabelling task of three classes based on grouped scenes. In this case we have mapped each of the ten scenes to one of:
\begin{enumerate}
\itemsep0em
\item Indoor (Airport, Indoor shopping mall, Metro station)
\item Outdoor (Pedestrian street, Public square, Street with Medium traffic, Urban park)
\item Transport (Tram, Bus, Metro)
\end{enumerate}
\vspace{-1.35em}
\paragraph*{Grouped scenes and city pairs}
The last of our three relabelling attempts combines the two previous ones using grouped scene types with the six cities for a new set of 18 classes consisting of all the grouped scene and city pairs.
\begin{figure}[t]
    \centering
    \includegraphics[width=\columnwidth]{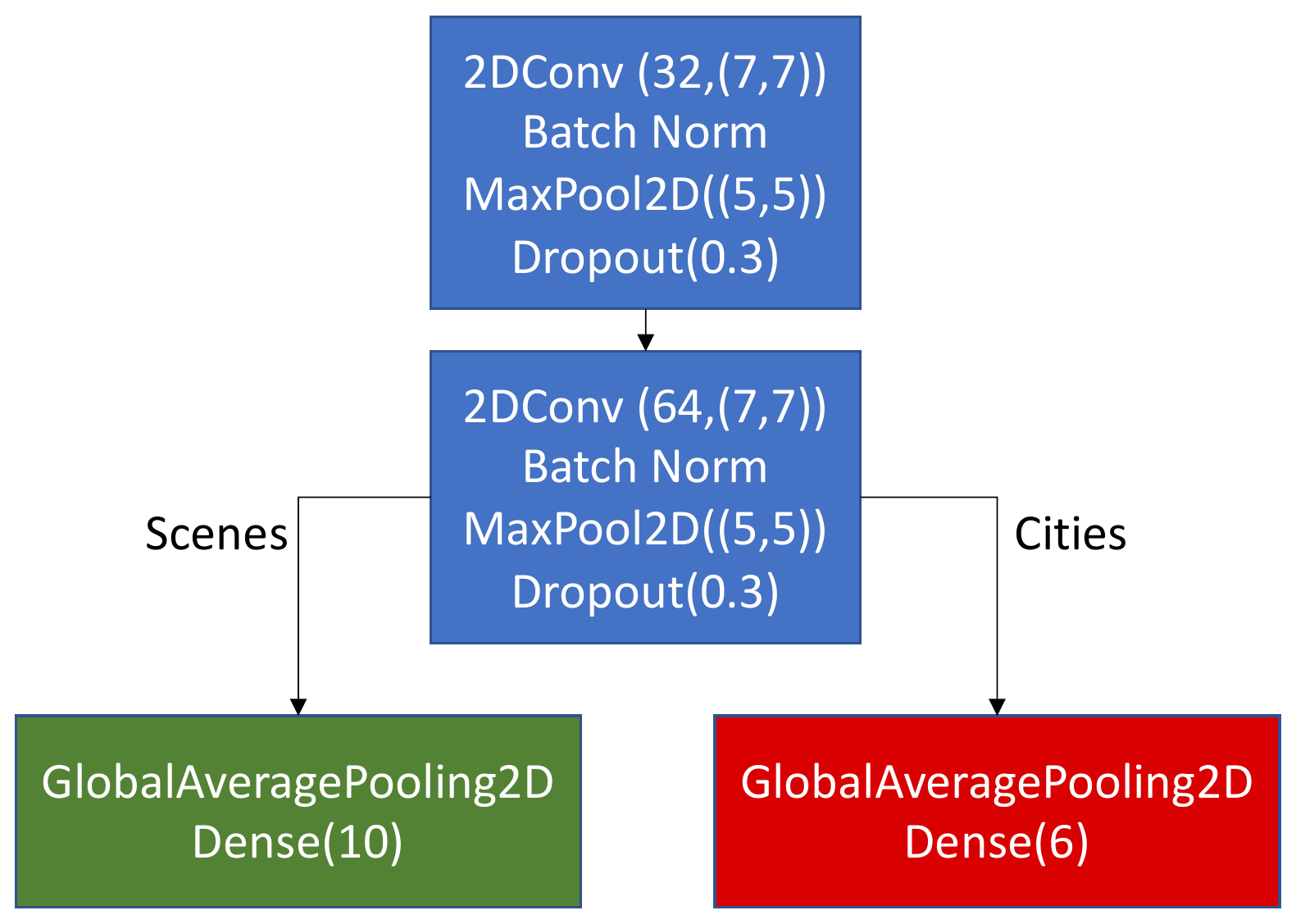}
    \caption{Multi-task learning model structure.}
    \label{fig:multi-task}
\end{figure}
\vspace{-1.35em}
\paragraph*{Multi-label classification}
Single-label, multi-class classification is training a model such that it makes one label prediction for each test sample where the labels is from a set of $N$ possible labels. Conversely, multi-label classification is the prediction of up to $N$ labels for each test sample \cite{cakir2015polyphonic}. This has been used before with joint-task work (see \cite{bear2019towards}) which strived to undertake ASC jointly with SED \cite{parascandolo2016recurrent}. The premise is such that instead of a complex model the problem is restructured as a 16-class problem: ten scene labels and six city labels. This produces a prediction matrix for all 16 classes from which we deduce three accuracy measures.
The matrix is spliced into two sub-matrices grouped by city and scene classes. The max for each test sample is taken as the prediction to calculate the accuracy for city and scene classes independently. The joint measure counts an accurate prediction when both the city and scene prediction is correct for a test sample. 
\vspace{-1.35em}
\paragraph*{Multi-Task classification}
Our final method is Multi-Task Learning (MTL), which has been used successfully in many domains and effectively learns two tasks simultaneously with one set of data \cite{morfi2018deep}. This is more successful when the tasks are closely related, such as in scene and city classification. MTL operates by sharing representations between related tasks, to generalize the model better on each task. 
Figure~\ref{fig:multi-task} shows the multi-task model structure we use to test city classification, which is simplified from \cite{morfi2018deep} to match our prior model. There are two outputs from one input. 
The model is very similar to our prior tests except for the two separate outputs. These represent the scene and city classification tasks respectively. For each task a loss function is optimised, each function is expressed as the weighted sum of both. We experimented to find the best weighting for both tasks and found $0.5$ (scenes) and $0.3$ (cities) to be optimal on the validation data. We note that whilst this does not sum to one, it is the same ratio between tasks as the ratio of classes in each task. With the optimal weights, we experimented training more epochs (300 and 400) with no significant increase in the prediction accuracy. 
We anticipate that MTL would improve city classification accuracy over multi-class city classification, but will not improve scene classification accuracy over single-label scene classification. This premise supports the cluster graph design taxonomy where labels can have a hierarchy but it is not enforced \cite{bear2018extensible}. 
\vspace{-0.3em}
\section{Analysis of Results}
All results are summarised in Table~\ref{tab:results} which begins with the DCASE ASC Subtask benchmark of 59.7\% accuracy. We discuss our results by the line order of this table. By adding one extra convolutional layer (line two in Table~\ref{tab:params}) we reduce the variation of each scene prediction for all scenes albeit with a minor $0.7\%$ decrease in overall accuracy. We retain the extra layer for a model robust to class variation as this is greater in the city problem. 

\label{sec:results}
\begin{table}[t]
    \centering
    \caption{Results from all approaches to city classification of soundscapes.}
    \begin{tabular}{l| r r r}
    Task & Accuracy & Task & nb\_classes\\
    \hline \hline
    Benchmark\cite{Mesaros2018_DCASE}  &  59.7\% & Scenes & 10 \\
    Extra layer CNN & 59\% & Scenes & 10 \\
    \hline
    Multi-class & 50\% & Cities & 6\\
    Scene priors & 26\% & Cities & 6 \\
    \hline
    Pairs & 37\% & both & 60\\
    Grouped & 89\% & Scenes & 3 \\
    Grouped pairs & 52\% & both & 18\\
    \hline
    Multi-label & 12\% & both & 16\\
    Multi-label & 32\% & Scenes & 10 \\
    Multi-label & 43\% & Cities & 6 \\
    \hline
    Multi-task & 57\% & Scenes & 10 \\
    Multi-task & \textbf{56\%} & Cities & 6\\
        
    \end{tabular}
    \label{tab:results}
\end{table}
\begin{table}[t]
\centering
\caption{City-wise classification accuracy with an CNN, and paired with grouped scenes.}

\begin{tabular}{l|r|r}
City & CNN Acc $\%$ & Paired Acc $\%$\\
\hline \hline
Barcelona     &  $40\%$ & $50\%$ \\
Helsinki     & $50\%$ & $40\%$ \\
London & $76\%$ & $47\%$\\
Paris & $32\%$ & $67\%$ \\
Stockholm & $81\%$ & $77\%$ \\
Vienna & $8\%$ & $27\%$ \\
\end{tabular}
\label{tab:results_city}
\end{table}
\begin{figure}[t]
    \centering
    \includegraphics[width=0.8\columnwidth]{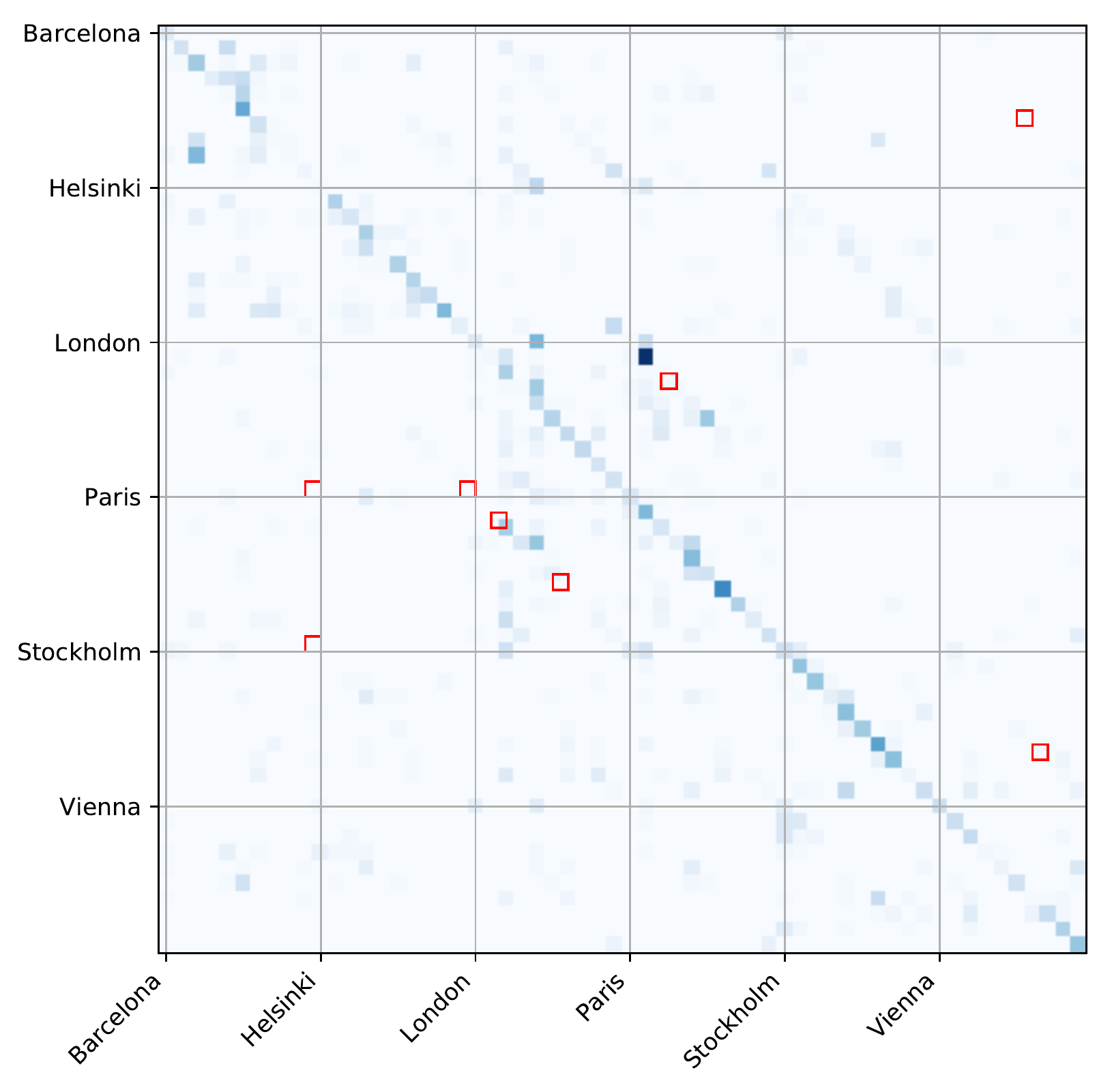}
    \caption{Confusions between 60 city-scenes.}
    \label{fig:city_scenesCM}
\end{figure}
\begin{table*}[!t]
\centering
\caption{City accuracy with ten scene-specific classifiers. This shows how cities vary by each scene.}
\begin{tabular}{l|rrrrrr|r}
Scene             & Barcelona    & Helsinki     & London       & Paris        & Stockholm    & Vienna       & Average      \\           
\hline \hline
airport           & 43.8\%        & 32.5\%        & 10.2\%        & 67.4\%        & 43.8\%        & 90.0\%        & 47.9\%            \\       
bus               & 13.9\%        & 0.0\%         & 2.5\%         & 44.7\%        & 20.8\%        & 0.0\%         & 13.7\%                   \\
metro             & 34.0\%        & 0.0\%         & 0.0\%         & 10.8\%        & 52.8\%        & 0.0\%         & 16.3\%              \\     
metro\_station     & 25.9\%        & 26.8\%        & 12.8\%        & 30.6\%        & 80.6\%        & 0.0\%         & 29.4\%               \\    
park              & 0.0\%         & 2.4\%         & 0.0\%         & 0.0\%         & 98.2\%        & 30.6\%        & 21.9\%             \\      
public\_square     & 72.2\%        & 0.0\%         & 44.4\%        & 0.0\%         & 30.6\%        & 0.0\%         & 24.5\%           \\        
shopping\_mall     & 5.6\%         & 75.0\%        & 10.1\%        & 100.0\%       & 52.8\%        & 2.8\%         & 41.0\%            \\       
street\_pedestrian & 32.5\%        & 0.0\%         & 11.1\%        & 5.6\%         & 83.3\%        & 29.6\%        & 27.0\%           \\        
street\_traffic    & 11.1\%        & 0.0\%         & 9.1\%         & 11.1\%        & 83.3\%        & 0.0\%         & 19.1\%           \\       
tram              & 60.9\%        & 0.0\%         & 1.8\%         & 22.6\%        & 41.4\%        & 0.0\%         & 21.1\%         \\          
\hline
Average           & 30.0\%        & 13.7\%        & 10.2\%        & 29.3\%        & 58.8\%        & 15.3\%        & 26.2\%\\
\hline
\end{tabular}
\label{tab:scenepriors}
\end{table*}

\begin{figure}[t]
    \centering
    \includegraphics[width=0.8\columnwidth]{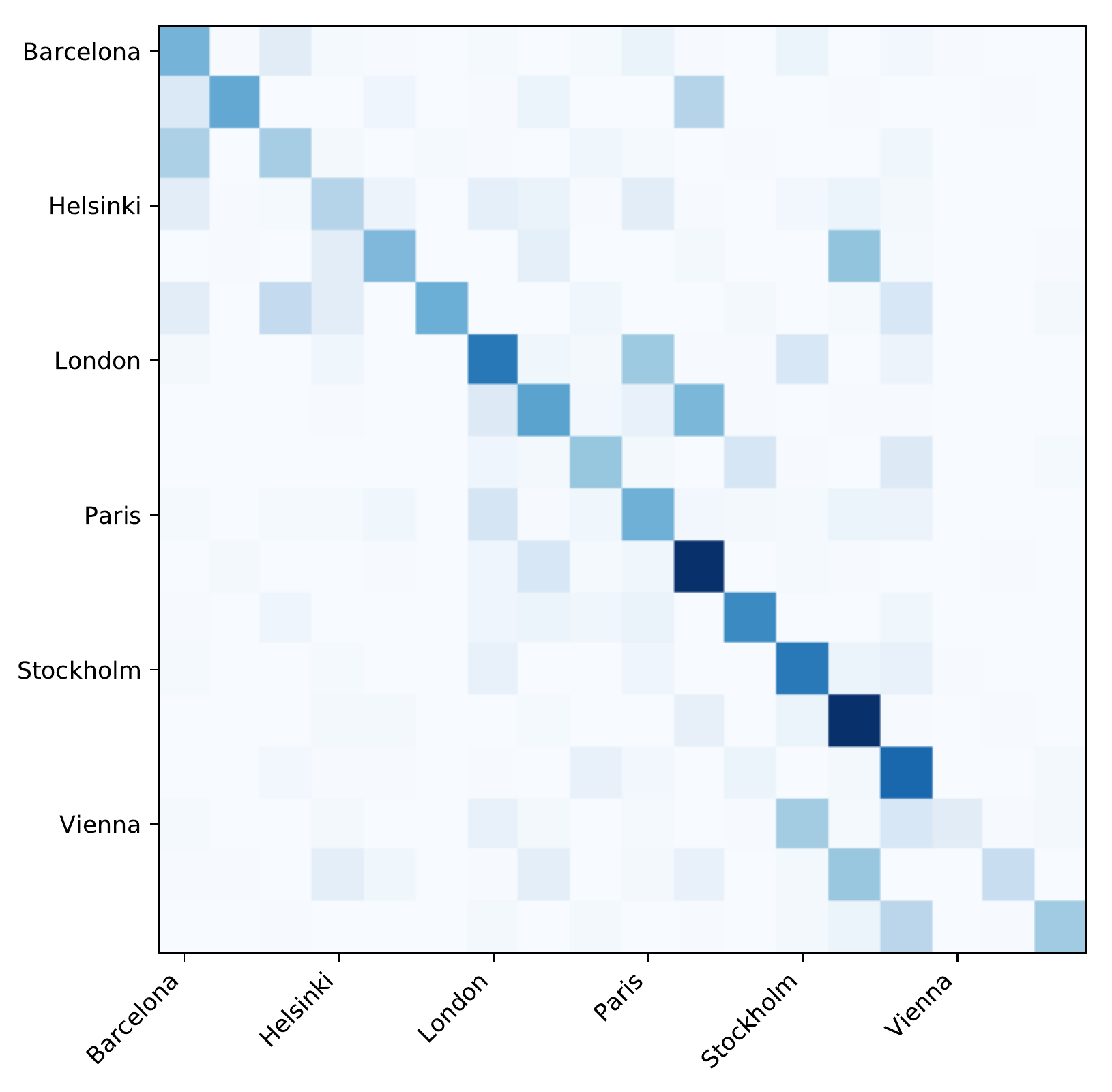}
    \caption{Confusions between 18 city-scenes, each city has three lines in `indoor', `outdoor', and `transport' order.}
    \label{fig:groupedcity_scenesCM}
\end{figure}
The multi-class city prediction with ten models previously trained on each of the ten scenes, scored a mean accuracy of $26\%$ (line four). These individual model accuracies are listed in Table~\ref{tab:scenepriors}. The scenes with lowest discrimination between cities are bus, metro, street\_traffic, and tram, suggesting that transport-based sound scenes have low variation irrespective of city. Airport scenes are the most distinctive for each city with an average accuracy of $47.9\%$ but this is biased towards Vienna suggesting Vienna airport has something very distinctive in its recordings. Similarly the park trained model appears to overfit to Stockholm. When comparing these results to the others in this paper, the lower number of training samples per classifier must have some effect and we attribute this to the lower average all classifier accuracy. 

The 60-class model accuracy is lower than the multi-class CNN, $37\%$ compared to $50\%$ (lines five and four in Table~\ref{tab:results} respectively). We use the 60-class model to produce a large confusion matrix to identify confusions between scenes and cities in Figure~\ref{fig:city_scenesCM} (Fig~\ref{fig:city_scenesCM} is online:\url{soundscape.eecs.qmul.ac.uk/misc-materials/} to zoom). The axes are labelled as the six cities, for each there are ten rows, each row representing the scenes in alphabetical order. 
With a lower accuracy it would be unwise to draw any confident conclusions; in Figure~\ref{fig:city_scenesCM} many city confusions are between the same scene. To show this, we take a greater than ten threshold to mean a noteworthy confusion, and highlight with red bounded squares in Figure~\ref{fig:city_scenesCM}. Confusions include: airport, metro, and street\_pedestrian. 
So while some scenes are possibly independent of city e.g. shopping\_mall, there do appear to be groups of scenes. This prompted the three-class problem based on grouped scenes. This model achieved $89\%$ accuracy (line six) and showed the majority of confusions are between indoor and outdoor scenes. It is possible that people/human actions are the common element causing this, and so transport based scenes are more discriminative.

Figure~\ref{fig:groupedcity_scenesCM} is a confusion matrix for the grouped-scenes and city pair labels. This model scored $52\%$ mean city accuracy (line seven in Table~\ref{tab:results}). By city results are in the third column of Table~\ref{tab:results_city} and show an improvement for three cities: Barcelona, Paris, and Vienna. Whilst the other three city accuracies are reduced, the model is more robust to all classes than the city-only label model (Table~\ref{tab:results}, line two). The confusions in Figure~\ref{fig:groupedcity_scenesCM} show that for all types of scenes Vienna is incorrectly predicted as Stockholm (not vice versa), suggesting there is still some bias towards Stockholm in this improved model. This is interesting given this pair of cities has the largest ED between their mean training data. The next greater confusions are between London and Paris, which has the lowest ED. In the future, better analysis of the feature spaces might help resolving these confusions to improve predictions. 

The penultimate test of multi-label classification scored a scene accuracy of $32\%$, city accuracy of $43\%$, and if a joint score of both must be correctly predicted, a joint accuracy of $12\%$ (Table~\ref{tab:results} lines eight, nine and ten respectively). These were scored by binarising the class predictions with a global threshold of $0.4$ and comparing the binarised matrix to the ground truth. Multiple thresholds were tested on the validation data before testing with $0.4$.

The final MTL approach scored $57\%$ for scenes ($2\%$ below the scene benchmark) and $56\%$ for cities (loss weights $0.5:0.3$ respectively), thus outperforming all other city classification methods with only a minor reduction in the scene task. Experiments to rebalance the loss weights can increase city but penalise scene classification. Results presented are best for both scene and city.
\vspace{-0.3em}
\section{Conclusions} \label{sec:conclusions}
In this paper we have addressed a novel task of classifying cities from sound scenes recorded in multiple locations and scenes in each city. We adapted prior data for a new task by relabelling. This work shows that adapting prior labelling schemes can increase city classification accuracy as per our grouped scene pair results. We attribute this observation to the greater variation within each city class in the feature space. This explains why our best result is achieved using the MTL method where learning scenes is a joint task with learning cities. Furthermore, we saw in our case that when the MTL weight ratio is similar to the class ratio for each task, the prediction for both tasks improved. 
Future work for city classification is vast; refining the models, augmenting the data, or analysing the inter-class variation of training data in the feature space.

\bibliographystyle{IEEEtran}
\bibliography{refs19}

\begin{thebibliography}{10}
\providecommand{\url}[1]{#1}
\def\UrlFont{\rmfamily}
\providecommand{\newblock}{\relax}
\providecommand{\bibinfo}[2]{#2}
\providecommand\BIBentrySTDinterwordspacing{\spaceskip=0pt\relax}
\providecommand\BIBentryALTinterwordstretchfactor{4}
\providecommand\BIBentryALTinterwordspacing{\spaceskip=\fontdimen2\font plus
\BIBentryALTinterwordstretchfactor\fontdimen3\font minus
  \fontdimen4\font\relax}
\providecommand\BIBforeignlanguage[2]{{%
\expandafter\ifx\csname l@#1\endcsname\relax
\typeout{** WARNING: IEEEtran.bst: No hyphenation pattern has been}%
\typeout{** loaded for the language `#1'. Using the pattern for}%
\typeout{** the default language instead.}%
\else
\language=\csname l@#1\endcsname
\fi
#2}}

\bibitem{virtanen2018computational}
T.~Virtanen, M.~D. Plumbley, and D.~Ellis, \emph{Computational analysis of
  sound scenes and events}.\hskip 1em plus 0.5em minus 0.4em\relax Springer,
  2018.

\bibitem{Guastavino2018}
\BIBentryALTinterwordspacing
C.~Guastavino, \emph{Everyday Sound Categorization}.\hskip 1em plus 0.5em minus
  0.4em\relax Springer International Publishing, 2018, pp. 183--213. [Online].
  Available: \url{https://doi.org/10.1007/978-3-319-63450-0_7}
\BIBentrySTDinterwordspacing

\bibitem{7545041}
B.~{Elizalde}, G.~{Chao}, M.~{Zeng}, and I.~{Lane}, ``City-identification of
  flickr videos using semantic acoustic features,'' in \emph{2016 IEEE Second
  International Conference on Multimedia Big Data (BigMM)}, April 2016, pp.
  303--306.

\bibitem{kumar2017audio}
A.~Kumar, B.~Elizalde, and B.~Raj, ``Audio content based geotagging in
  multimedia,'' \emph{Proc. Interspeech 2017}, pp. 1874--1878, 2017.

\bibitem{1416352}
R.~G. {Malkin} and A.~{Waibel}, ``Classifying user environment for mobile
  applications using linear autoencoding of ambient audio,'' in \emph{Proc'
  IEEE International Conference on Acoustics, Speech, and Signal Processing
  (ICASSP)}, vol.~5, March 2005, pp. v/509--v/512 Vol. 5.

\bibitem{4036742}
S.~{Chu}, S.~{Narayanan}, C.~.~J. {Kuo}, and M.~J. {Mataric}, ``Where am i?
  scene recognition for mobile robots using audio features,'' in \emph{2006
  IEEE International Conference on Multimedia and Expo}, July 2006, pp.
  885--888.

\bibitem{bear2018extensible}
H.~Bear and E.~Benetos, ``An extensible cluster-graph taxonomy for open set
  sound scene analysis,'' in \emph{Workshop on Detection and Classification of
  Acoustic Scenes and Events}, 2018.

\bibitem{Mesaros2018_DCASE}
A.~Mesaros, T.~Heittola, and T.~Virtanen, ``A multi-device dataset for urban
  acoustic scene classification,'' in \emph{arXiv preprint arXiv:1807.09840},
  November 2018, pp. 9--13.

\bibitem{bisot2017feature}
V.~Bisot, R.~Serizel, S.~Essid, and G.~Richard, ``Feature learning with matrix
  factorization applied to acoustic scene classification,'' \emph{IEEE/ACM
  Transactions on Audio, Speech, and Language Processing}, vol.~25, no.~6, pp.
  1216--1229, 2017.

\bibitem{cakir2015polyphonic}
E.~Cakir, T.~Heittola, H.~Huttunen, and T.~Virtanen, ``Polyphonic sound event
  detection using multi label deep neural networks,'' in \emph{2015
  international joint conference on neural networks (IJCNN)}.\hskip 1em plus
  0.5em minus 0.4em\relax IEEE, 2015, pp. 1--7.

\bibitem{bear2019towards}
H.~Bear, I.~Nolasco, and E.~Benetos, ``Towards joint sound scene and polyphonic
  sound event recognition,'' in \emph{Arxiv.org
  \url{https://arxiv.org/abs/1904.10408v1}}, 2019.

\bibitem{parascandolo2016recurrent}
G.~Parascandolo, H.~Huttunen, and T.~Virtanen, ``Recurrent neural networks for
  polyphonic sound event detection in real life recordings,'' in \emph{2016
  IEEE International Conference on Acoustics, Speech and Signal Processing
  (ICASSP)}.\hskip 1em plus 0.5em minus 0.4em\relax IEEE, 2016, pp. 6440--6444.

\bibitem{morfi2018deep}
V.~Morfi and D.~Stowell, ``Deep learning for audio event detection and tagging
  on low-resource datasets,'' \emph{Applied Sciences}, vol.~8, no.~8, p. 1397,
  2018.

\end{thebibliography}

\end{sloppy}
\end{document}